\documentclass[aps,prb,twocolumn,showpacs,superscriptaddress,letterpaper]{revtex4-1}
\usepackage{graphicx}
\usepackage{epsfig}
\usepackage{subfigure}
\usepackage{times}
\usepackage{color}
\usepackage{epstopdf}
\usepackage{amsmath,amsthm,amssymb}
\usepackage{bm}

\begin{document}

\title{Pulse and quench induced dynamical phase transition in a chiral multiferroic spin chain}
\author{M.~Azimi}
\affiliation{Institut f\"ur Physik, Martin-Luther-Universit\"at Halle-Wittenberg, D-06099 Halle, Germany}
\author{M.~Sekania}
\affiliation{Center for Electronic Correlations and Magnetism, Institute of Physics, University of Augsburg, D-86135 Augsburg, Germany}
\affiliation{Andronikashvili Institute of Physics, Tamarashvili~6, 0177 Tbilisi, Georgia}
\author{S.~K.~Mishra}
\affiliation{Department of Physics, Indian Institute of Technology, Banaras Hindu University, Varanasi 221005, India}
\author{L.~Chotorlishvili}
\affiliation{Institut f\"ur Physik, Martin-Luther-Universit\"at Halle-Wittenberg, D-06099 Halle, Germany}
\author{Z.~Toklikishvili}
\affiliation{Department of Physics, Tbilisi State University, Chavchavadze av.~3 0128, Tbilisi, Georgia}
\author{J.~Berakdar}
\affiliation{Institut f\"ur Physik, Martin-Luther-Universit\"at Halle-Wittenberg, D-06099 Halle, Germany}

\begin{abstract}
Quantum dynamics of  magnetic order in a chiral  multiferroic chain is studied.
We consider two different scenarios:
Ultrashort terahertz (THz) excitations or a sudden electric field quench.
Performing analytical
and numerical exact diagonalization calculations
we trace the pulse induced spin dynamics and extract quantities that are relevant to quantum information processing.
In particular, we analyze the dynamics of  the system chirality, the von~Neumann entropy, the pairwise and the many body entanglement.
If the characteristic frequencies of the generated states are non-commensurate then a partial loss of pair concurrence occurs.
Increasing  the system size  this effect becomes even more pronounced. Many particle entanglement and chirality are robust and persist  in the incommensurate phase.
To analyze the dynamical quantum transitions for the quenched and pulsed dynamics we combined the Weierstrass factorization technique for entire functions and Lanczos exact diagonalization method.
For a small system we obtained analytical results including the rate function of  Loschmidt echo. Exact numerical calculations for a system up to  $40$ spins confirm phase transition. Quench-induced dynamical transitions have been extensively studied recently. Here we show that related dynamical transitions can be achieved and controlled by appropriate electric field pulses.

\end{abstract}
\pacs{}

\date{\today}

\maketitle

\section{\label{sec:intro}Introduction}

Multiferroic (MF) materials and composites possess simultaneously a multiple of primary ferroic orderings that are possibly coupled.
\cite{Schmid,Khomskii,Ramesh,Nan,Scott,Azimi1,Katsura,Cheong,Park, Menzel,Azimi2,Malashevich,Sneff,Walker,Staruch}.
The intense research interest in these materials is fueled by a multitude of potential applications
as well as by the possibility of employing them as a testing ground for fundamental questions concerning the interplay between
magnetism, electricity, electronic correlations, and symmetries. For example MF magnetoelectrics allow the  control of the magnetic order by an external electric field  (in addition to a magnetic field)
by virtue of the magnetoelectric (ME) coupling which renders new concept for data storage and read/write schemes, potentially    at low-energy consumption.
MF have a long history \cite{Schmid}.
Their utilization were however hampered  by the notoriously weak MF coupling.
Novel nano fabrication and characterization techniques, especially for heterostructures  with stronger and controlled MF coupling
gave a new impetus to the field with numerous findings and applications (see e.g., Refs.~[\onlinecite{spaldin,Eerenstein}] and references therein). The underlying mechanisms for MF coupling are diverse.
Of special interest here is the (spin-dependent) charge driven ME coupling in non-collinear magnetic compounds (cf. Ref.~[\onlinecite{Tokura}] and references therein).
For example, the perovskite type manganites RMnO$_3$ (R = Tb, Dy, Eu$_{1-x}$Y$_x$)
show in a certain temperature range a transversal helical (cycloidal) spin order and may exhibit a ferroelectric polarization which
is determined by the topology of the spins \cite{35,36,37}.
We will explore a particular issue, namely the functionalization of these MF materials and their special ME coupling to
time-dependent quantum information processing via electric THz pluses.
In this context, we mention previous studies \cite{Azimi1,Azimi2} employing a \emph{static} external  electric field that were shown to
enhance the quantum state transfer fidelity \cite{Azimi2} or/and  the increase in efficiency of  quantum heat engines
\cite{Azimi1} based on spiral multiferroic working substance.
The quantum dynamical response of these materials and a possible coherent control of the response, e.g. with E-field pulses, was
not studied to our knowledge and is the topic of this study.
In particular we are interested
in the possibility of triggering dynamical quantum transition via external fields.
In the vicinity of a dynamical quantum phase transition a system with an Hamiltonian for which the eigenvalues depend analytically on the parameters
may exhibit a non-analytic behavior when approaching the thermodynamic limit.
A relevant experimental setup is a quantum quench, i.e. a sudden change of the driving parameter in the Hamiltonian, or an electromagnetic pulse (see below).
Recently the connection between the singularities in the nonequilibrium dynamics of a quantum system and the theory of equilibrium phase transitions was discussed \cite{heyl}.
Relevant issue is  the well-established  theorem of Lee and Yang \cite{Yang} concerning the
connection of the zeros of the equilibrium partition function when continued to  complex conjugate fields or to the complex temperature plane.
We refer also to the pioneering works of S.~Grossmann, W.~Rosenhauer and M.~E.~Fisher \cite{Grossmann}. Experimental observations of the Lee-Yang zeros was reported in Ref.~[\onlinecite{Peng}].
In the context of this work we refer to Refs.\cite{heyl,karrasch,vajna,hickey}
where it has been argued that the dynamical quantum phase transitions are indicated by changes in the zeros
of rate function of the return probability of the Loschmidt echo.

In order to identify true quantum phase transitions and separate them from  resonance related phenomena the robustness of the transition with respect to the system's size should be checked.
Furthermore, also in the equilibrium case,  the term ``phase transition''  in a finite  system
refers to  precursors of a real phase transitions occurring in the thermodynamic limit.
Yet, this criterion is not universal as it applies only to cases where such a thermodynamic limit really exists.
In some cases achieving  the thermodynamic limit is not possible at all.
A  typical example are sodium clusters \cite{Ellert} or ferrofluid clusters \cite{Borrmann1}.
The same behavior occurs in other complex finite systems as well, for example for  atomic nuclei \cite{Cejnar}.
In these materials the nature of  phase transitions changes with increasing the system's size.
Therefore, achieving the thermodynamic limit for  certain systems
which do not have universal scaling properties is questionable.
Phase transitions in  finite systems have been the subject of numerous recent works,
e.g. Refs.~[\onlinecite{Borrmann2,Aslyamov, Hernandez-Rojas, Hilbert, Cejnar2, Tarasov}].
Dynamical phase transitions in the thermodynamic limit can be detected  using transfer matrix approach \cite{Andraschko1}.
A combination of the above aspects will be relevant for the present project,
as we will be interested in quantum transitions in a field driven, chiral MF system.
The so-called quench dynamic, i.e. a sudden change in the Hamiltonian governing the system has been recently in the focus of research \cite{heyl,karrasch,vajna,hickey}.
The present work shows that exposing the system to  short THz pulses, that are already feasible, result in similar collective transition phenomena.
The pulses offer in addition the possibility to coherently control the pathways for these transitions.

Special attention is devoted to  a specific chiral spin order formation and its dependence on the external field.
The competition  between the spin exchange and  the spin-orbital coupling (Dzyloshinski-Moriya interactions) that
is naturally rooted in these compounds is the source for the formation of the chiral ground state.
The emergent electric polarization ${\vec{P}}$  is akin to the non-collinear spin order as it is proportional to $\hat{e}_{i,i+1}\times(\vec{S}_{i}\times\vec{S}_{i+1})$,  and hence ${\vec{P}}$ disappears in the collinear case of a
fully aligned ferromagnetic ground state.
Here  $\hat{e}_{i,i+1}$ is the unit vector connecting neighboring spins.
The vector spin chirality $\vec{{\bm{\kappa}}}=\langle\vec{S}_{i}\times\vec{S}_{i+1}\rangle$ is a quantitative characteristic measure of the chiral state.
If the chirality is zero, $\vec{\bm{\kappa}}=0$, the system turns  insensitive to an electric field.
Switching of a magnetic order by an optical pulse  reduces significantly the time scale required
for a quantum information processing.
On the other hand an optical pulse may lead to a strong structural change in the MF ground state
and eventually to a lose of quantum coherence, which is a vital ingredient for  quantum information processing \cite{Azimi2}.
The robustness of the ground state  upon the application  of an optical pulse is another issue of concern.
E.g., in the experiments of Refs.~[\onlinecite{Kubacka,Johnson}] the nonequilibrium
increase in magnetic disorder were induced by femtosecond laser pulse.
After applying a pulse in the range of $400$~femtoseconds (fs) to $2$~picoseconds (ps) a transition
from the commensurate phase to a spiral incommensurate phase was observed \cite{Johnson}.
{By commensurate/incommensurate transition we refer to not only  the geometry of the non-collinear spin configuration formed after a quench (pulse), but as well to the characteristic frequencies of the states $\omega_{n}= (E_{n} - E_{0})/\hbar$ involved in the dynamics ($E_{0}$ is the ground state).  We are particularly   interested in the latter aspect of the  commensurability of the characteristic frequencies. We note in this context that in a finite quantum system,  quantum revivals are closely related to the commensurability between different characteristic frequencies of the system \cite{Schleich}. If the system after the pulse (quench) turns into an incommensurate frequency phase a periodic or a quasi-periodic quantum revival in chirality are not expected.}

Chirality and entanglement are interrelated and chirality can be considered as a witness of quantum entanglement \cite{pachos}.
The model under study fulfills both requirements: the system can be manipulated by optical pulse and by a sudden quench.
We will inspect the chiral state formed in the system after the action of  pulse excitations or a sudden quench by studying a periodic in time quantum revivals in the chirality and entanglement.
Our model is relevant for the one phase chiral MF, e.g., ${\rm LiCu}_2{\rm O}_2$, ${\rm CoCr}_2{\rm O}_4$, ${\rm LiVCuO}_4$  \cite{Yamasaki}.
Depending on the system size we were able to perform full analytical calculations for the time propagation  and extract some trends for the entanglement and chirality evolution and to relate that to the underlying physics.
For larger systems we resort to full numerical exact diagonalization methods.
For more insight we will inspect equal-time spin-spin and chirality-chirality correlation functions.
In particular our interest concerns chirality, a
quantity which is an order parameter for one phase MF system and is related to the ferroelectric polarization of the MF systems.
Any remarkable change in chirality that might occur during the dynamical phase transition is of high experimental relevance.

The paper is organized as follows:
Section~\ref{sec:Model} introduces the model and basic notation,
Section~\ref{sec:NumSim} presents results of the numerical calculations for several entanglement witnesses such as: concurrence, von~Neumann entropy, one and two tangle. In  section~\ref{sec:DynamicalPT} we study the  dynamical quantum  transitions.
Using Weierstrass factorization technique we will study the zeros of the rate function of the Loschmidt echo.
We also inspect the vector chiral and nematic phases of the system and the spin-spin correlation functions.

\section{\label{sec:Model}Theoretical foundation}
We consider a one-dimensional  quantum spin chain along the $x$ axis  with a charge-driven multiferroicity.
The chain is subjected to pulses of an electric field  $E(t)$ that  is linearly polarized along the $y$ axis.
Additionally, an external magnetic field $B$ is applied along the $z$ axis. This coordinate system applies both to the spin and charge dynamics.
A  Hamiltonian capturing this situation reads
\begin{eqnarray}\label{Hamiltonian1}
 \hat{H}&=&J_1\displaystyle\sum_{i=1}^L\vec{S}_i\cdot\vec{S}_{i+1}+J_2\displaystyle\sum_{i=1}^L\vec{S}_i\cdot\vec{S}_{i+2}\nonumber\\
        &-&B\displaystyle\sum_{i=1}^LS_{i}^{z}+E(t)g^{\phantom{+}}_{\mathrm{ME}}\sum_{i}^L(\vec{S}_{i}\times\vec{S}_{i+1})^z.
\end{eqnarray}
The nearest neighbor exchange coupling  of a  spin $1/2$ (denoted $\vec{S}_i$) localized at site $i$ is ferromagnetic $J_1<0$, while next-nearest neighbor
interaction is antiferromagnetic $J_2>0$.
We consider periodic boundary conditions such that $\vec{S}_{L+i}=\vec{S}_i$.
The time dependent electric field $E(t)$ couples to the electric polarization as
$-\vec{E}(t)\cdot\vec{P}=E(t)g_{\mathrm{ME}} \sum_{i=1}^L(\vec{S}_{i}\times\vec{S}_{i+1})^z$,
where $g^{\phantom{+}}_{\mathrm{ME}}$ is the magnetoelectric coupling strength.
The quantity $\kappa_i=(\vec{S}_{i}\times\vec{S}_{i+1})^z$ is the $z$ (longitudinal) component of the vector chirality (VC).
In absences of frustration ($J_2=0$), i.e. for co-linear spin order, $\kappa_i$ and $\vec{P}$ vanish, and the chain does not react to $\vec{E}(t)$.
The ground-state and the dynamical properties of the parent Hamiltonian systems ($g_\mathrm{ME}=0$),
the so called frustrated ferromagnetic spin $\frac{1}{2}$ Heisenberg chains,
have been widely discussed in the  literature \cite{Chubukov2001,Sirker1}.
The phase diagram for a wide range of parameters and net magnetization have been studied \cite{Chubukov2001},
and was found to exhibit a rich variety of phases, including vector chiral phase, the nematic phase, and other multipolar phases.
Here we will focus on the {\em time-dependent dynamics} of the system with the spin-orbital coupling ($g_\mathrm{ME}\neq0$)
in the vicinity of the saturated magnetization for the fixed ratio of the spin-exchange couplings.

Each term of the Hamiltonian $\hat{H}$ (Eq.~\eqref{Hamiltonian1}) commutes with the total $z$-component of the spin, $S^z=\sum S_i^z$.
Hence $S^z$ is a good quantum number and any eigenstate of the system can be characterized by the total number of `down' (or `up') spins in the studied system.
In this manuscript we adopt the term ``$n$-excitations'' in order to denote the number of $n$ `down' spins.
For instance $n$-excitations spin state means $S^z=S-n$ (in  units of $\hbar$), where $S=L/2$.
In the case of only ferromagnetic interaction,
zero excitation spin state corresponds to a fully aligned (saturated) ferromagnetic ground state of the system.

The effective model Hamiltonian (Eq.~\eqref{Hamiltonian1}) is relevant for 1D spin frustrated MF oxides, e.g. LiCu$_{2}$O$_2$.
Typical values of the relevant parameters are  $-J_{1}\approx J_{2}$;
e.g. for LiCu$_{2}$O$_2$ one finds \cite{Park}
$J_{1}\approx -11 \pm 3 $ meV and $J_{2}\approx 7 \pm 1 $ meV.
For clarity  we adopt dimensionless units and measure energies  in units of $J_2$. The time is measured in units of $\frac{\hbar}{J_{2}}$ which typically in the range of $\approx0.5$[ps].

For coherent control of the system on a ps time scale, we
apply terahertz (THz) electric field pulses with a DC component, i.e. we use an electric field
$E(t)=E_0+E_1(t)$, where $E_0$ and $E_1(t)$ are  respectively static and time-dependent electric fields,  both linearly polarized along the $y$ axis.
The Hamiltonian of the driven system can be written as
\begin{eqnarray}\label{Hamiltonian2}
 \hat{H}=\hat{H_0}+\hat{H_1},
\end{eqnarray}
where $\hat{H_0}$ is the unperturbed Hamiltonian with the static electric field $E_0$
\begin{eqnarray}\label{ham0}
 \hat{H}_0&=&J_1\displaystyle\sum_{i=1}^L\vec{S}_i\cdot\vec{S}_{i+1}+J_2\displaystyle\sum_{i=1}^L\vec{S}_i\cdot\vec{S}_{i+2}
        -B\displaystyle\sum_{i=1}^LS_{i}^{z}\nonumber\\
        &+&E_0 g^{\phantom{+}}_{\mathrm{ME}}\sum_{i}^L(\vec{S}_{i}\times\vec{S}_{i+1})^z
\end{eqnarray}
and
\begin{eqnarray}
 \hat{H_1}=E_1(t)g^{\phantom{+}}_{\mathrm{ME}}\sum_{i}^L(\vec{S}_{i}\times\vec{S}_{i+1})^z.
\end{eqnarray}
The system is subjected  at $t = 0$ to a pulse of a rectangular shape, i.e.
\begin{eqnarray}\label{Pulse}
 E_1(t) = \begin{cases} \frac{E_{1}^{0}}{\varepsilon} & \text{if} -\frac{\varepsilon}{2} < t < \frac{\varepsilon}{2}  \\
 0 & \text{otherwise}  \end{cases}.
\end{eqnarray}
The width of the applied pulse is $\varepsilon \approx 0.3$  which translates in SI units to $\approx 0.15$[ps].

For  small $\varepsilon$, it is reminiscent to a $\delta$-kick with the strength $E_{1}^{0}$.
Small $\varepsilon$ means here that the pulse is shorter than the transition times between those states of the spectrum  that are involved in the dynamics.
Hence, we can assess this condition a posterior, i.e. after having diagonalized the Hamiltonian and figured out the highest pulse excited levels. It is worth noting that such pulses can be realized as a sequence of  highly asymmetric (in time) propagating single cycle THz pulses.
The one strong and short half cycle acts as the ``pulse" in eq.~\eqref{Pulse} and the second long and flat half cycle serves
as the DC field (cf. \cite{arxiv} for more details).\\
Hereafter we abbreviate $d^{\phantom{0}}_1=E^{0}_{1}g^{\phantom{+}}_{\mathrm{ME}}, \quad d^{\phantom{0}}_0=E^{\phantom{0}}_0g^{\phantom{+}}_{\mathrm{ME}}.$
The time evolution of the system is given by the Schr\"{o}dinger equation
\begin{eqnarray}\label{Schrdinger equation}
 i\frac{\partial}{\partial t}\vert\psi(t)\rangle=\hat{H}\vert\psi(t)\rangle,
\end{eqnarray}
with the initial condition $\vert\psi(t=-\frac{\varepsilon}{2})\rangle=\vert\psi_0\rangle$.
Here $\vert\psi_0\rangle$ is the initial state of the system ($\hat{H}_0$), i.e. before applying the pulse.
The pulse-triggered coherent state propagates under the influence of $\hat{H_0}$ for $t>\frac{\varepsilon}{2}$.
After rescaling the time $T=(t+\frac{\varepsilon}{2})/\varepsilon$,
($t=-\frac{\varepsilon}{2}$, $T=0$; $t=\frac{\varepsilon}{2}$, $T=1$) and
within the short-pulse assumption, as explained above, we find
the state of the system right after the pulse to be  governed by the Magnus-type propagation \cite{arxiv}
\begin{eqnarray}\label{State1}
 \vert\psi(T=1)\rangle=e^{-i\hat O}\vert\psi(T=0)\rangle,
\end{eqnarray}
where the operator ${\hat O}$ is given by
\begin{equation}\label{eq:d1}
 \hat O=d_1\sum_{i=1}^L(\vec{S}_{i}\times\vec{S}_{i+1})^z.
\end{equation}
We note, that expression (7) includes all non-linear terms in the electric fields strength (as long as the short-pulse approximation is viable \cite{arxiv}) and can deal thus with intense-field dynamics.
After the pulse the system evolves  in time under the action of the unperturbed
Hamiltonian $\hat{H_0}$. Thus,  the state at any arbitrary time $t'$ after the pulse can be written as
\begin{eqnarray}\label{State3}
 \vert\psi(t')\rangle=\sum_{n}e^{-i\mathcal{E}_n t^\prime }\vert\phi_n\rangle \langle\phi_n\vert e^{-i\hat O}\vert\psi(0)\rangle,
\end{eqnarray}
where $\vert\phi_n\rangle$ and $\mathcal{E}_n$ are the eigenstates and corresponding eigenvalues of the Hamiltonian $\hat H_0$ (Eq.~(\ref{ham0})).
$\vert\psi(0)\rangle$ is the initial state in which the system is prepared right before applying the pulse (time $t=0$).
This state, in general, can be chosen to be any coherent state of the system, not only the ground state.
Obviously,
the procedure can be repeated stroboscopically if a further pulse is applied at some $t'$.
In this case $\vert\psi(t')\rangle$ will replace $\vert\psi(0)\rangle$ in Eq.~\eqref{State3}. The propagation scheme does not allow for any insight in what happens during the pulse.

Alternatively we also consider scenario when we suddenly quench the electric field.
Here we again consider the Hamiltonian (\ref{Hamiltonian2}),
but the protocol of the time evolution is as follows:
for $t<0$ the system is given by $\hat{H}=\hat{H_0}+\hat{H_1}$
and at $t=0$ it is suddenly quenched to $\hat{H}=\hat{H}_0$.
Hence, after the quench, the system evolves in time again   under the influence of the unperturbed Hamiltonian $\hat{H_0}$
and the time-evolved state at any arbitrary time $t'>0$ is given as
\begin{eqnarray}\label{State4}
 \vert\psi(t')\rangle=\sum_{n}e^{-i\mathcal{E}_n t^\prime }\vert\phi_n\rangle \langle\phi_n\vert\psi_0\rangle,
\end{eqnarray}
but here $\vert\psi_0 \rangle$ denotes the ground state of the full Hamiltonian $\hat{H}$ in Eq.~\eqref{Hamiltonian2}.

\section{\label{sec:NumSim} Pulse induced dynamics of chirality and entanglement}

In this section we analyze the system-size effect on the chirality,
two-tangle and von~Neumann Entropy.
We will mainly focus on the results obtained by exact numerical diagonalization and time evolution,
like Lanczos algorithm or some other Krylov-subspace based methods \cite{Simoncini}.
To introduce the terminology illustrated by analytical expressions we also consider a system of 4 spins.
We start with the time evolution of the chirality.

If the system is kept initially in a one-excitation ground state
\begin{figure}[!t]
 \includegraphics*[width=.45\textwidth]{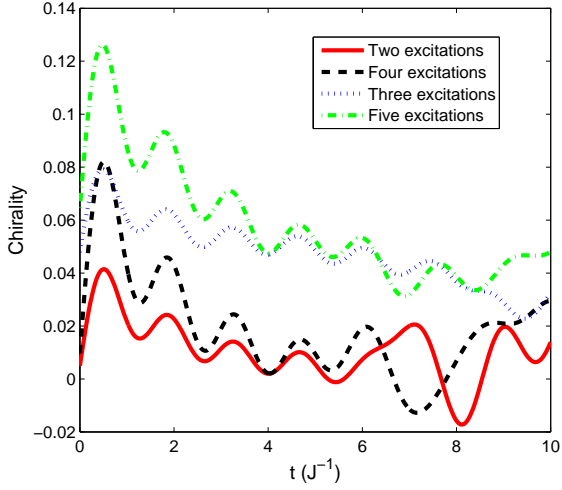}
 \caption{Chirality $\kappa_i$ as a function of the time  for two-,
          three-, four- and five-excitations initial state for $L=30$ spin chain with periodic boundary conditions.
          The duration of the applied pulse is $\varepsilon \approx 0.3$ and its  strength is $d_1=0.5$.
          The spin-exchange couplings and the initial electric-field strength are $J_1=-J_2=-1.0$ and $d_0=0.05$, respectively.
          Note that $J_2$ sets the energy units.
         }
 \label{chiral_L20}
\end{figure}
\begin{figure}[!t]
\includegraphics*[width=0.45\textwidth]{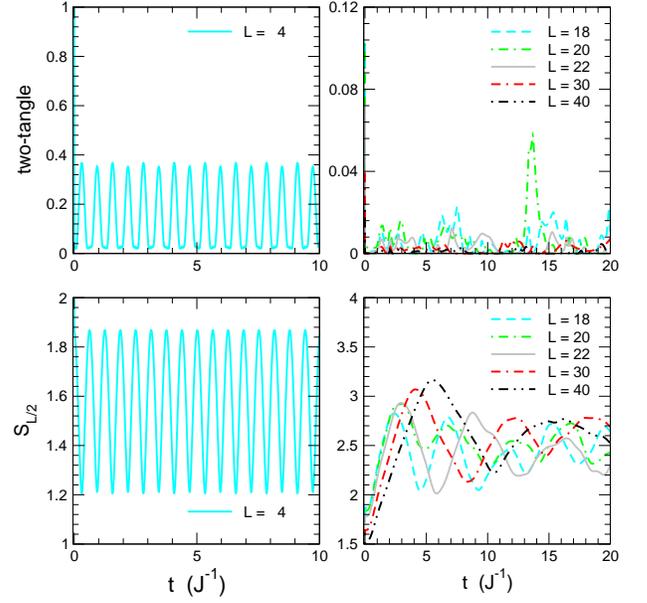}
\caption{Time dependence of the averaged two tangle (upper raw) and the von~Neumann entropy (lower raw)
         for $L = 18, 20, 22, 30, 40$ (right column) and $L=4$ (left column) chain sizes.
         In all the cases the system is in a two-excitation ground state.
         The duration of the applied pulse is $\varepsilon \approx 0.3$ and the strength $d_1=0.5$.
         The spin-exchange couplings and initial electric-field strength are $J_1=-J_2=-1.0$ and $d_0=0.05$, respectively.
        }
\label{av_tangle_ent}
\end{figure}
the sum of expectation values of the two terms $\langle S_i^-S_{i+1}^+\rangle$ and $\langle S_i^+S_{i+1}^-\rangle$
add up to the same constant for any time $t$, irrespective of chain lengths.
In the case of higher $n$-excitations ground states, however,
the chirality shows oscillations with time.
The oscillatory behavior of chirality varies with the choice of initial state.
As shown in Fig.~\ref{chiral_L20} the peak of the chirality increases with higher $n$-excitation ground states.
Also the even(odd)-excitations initial states follow a similar pattern
of oscillations which differs from odd(even)-excitations initial state case.

To quantify the entanglement we use one- and two-tangle as a measure of
non-local and local correlations in the system\cite{Wooters,Lakshminarayan}.
One-tangle is given by $\tau_1=4{\rm det }\rho_1$, where $\rho_1$ is the reduced density matrix for
a single spin after tracing out the rest, and two-tangle reads  $\tau_2=\sum_{m=1}^{N}C_{nm}^{2}$, where $N$ is the number of spins,
$C_{nm}$ is the pair concurrence between spins $n$ and $m$ in the system, defined as
\begin{eqnarray}\label{Concurrence}
 C_{nm}={\rm max}(0,\sqrt{R_1}-\sqrt{R_2}-\sqrt{R_3}-\sqrt{R_4}).
\end{eqnarray}
$R_n$ are the eigenvalues of the matrix
$R=\rho_{nm}^{R}(\sigma_{1}^{y}\bigotimes\sigma_{2}^{y})
(\rho_{nm}^{R})^{*}(\sigma_{1}^{y}\bigotimes\sigma_{2}^{y})$ and $\rho_{nm}^{R}$ is the
reduced density matrix of the system
obtained from the density matrix $\hat{\rho}=|\psi(t)\rangle\langle\psi(t)|$ after
retaining spins at $n$ and $m$ positions and tracing out the rest.
For the four spin system with ground states $\vert\phi_2\rangle$ or $\vert\phi_7\rangle$ (see Appendix~\ref{sec:appA}), one tangle
$\tau_1$ is unity and is independent of time, therefore it is less interesting for us.
The reason why $\tau_1$ does not vary with time follows from the structure of the single qubit reduced density matrix.
The off-diagonal terms $\langle S_k^+\rangle$ and $\langle S_k^-\rangle$ vanish for $|\phi_2\rangle$ and $|\phi_7\rangle$ states
(no more excitation is permitted for a fixed value of $d_0$ and $B$).
Hence, the determinant of the reduced density matrix $\rho_1$ which reads $\frac{1}{4}-\langle S_1^z\rangle^2$ is constant.

The two-tangle  in the case of the two-excitation ground state $|\phi_7\rangle$,
oscillates with time (see Fig.~\ref{av_tangle_ent}).
We note that this similarity in the behaviors of chirality and local entanglement (two-tangle)
is akin to  small systems.
For larger systems, the chirality shows a more elaborate time evolution as compared to $L=4$.

This nonequilibrium oscillation in chirality goes along with an oscillation of the emergent electric
polarization and hence might be detected experimentally either by a time dependent electric susceptibility measurement
or by detecting the emitted radiation.

In Fig.~\ref{av_tangle_ent} we show the time evolution of the averaged two-tangle for various sizes of the spin chain in the two-excitation ground state.
In all shown cases the oscillatory pattern is visible.
We observe, however, that the two-tangle decreases significantly with the increasing chain size in contrast to the case of chirality. For $L\geqslant 18$ the two-tangle almost disappears during the  pulse-free time evolution.
Thus local entanglement is less robust and does not survive after significant pulse induced  changes in the system.

Another measure of the entanglement, which quantifies better the multiparticle entanglement, is the von~Neumann entropy.
For a system of $L$ spins the von~Neumann entropy of the bipartition can be defined as
\begin{eqnarray}\label{vnent}
 S_{L/2}=-{\rm Tr}_{1,\dots,L/2}[\rho_{1,\dots,L/2}\log_2(\rho_{1,\dots,L/2})],
\end{eqnarray}
where, the reduced density matrix of the first $L/2$ spins is given by
$ \rho_{1,\dots,L/2}={\rm Tr}_{L/2+1,\dots,L}\left(\rho_{1,\dots,L}\right)$.

%

Fig.~\ref{av_tangle_ent}  shows the time evolution of von~Neumann entropy for various chain sizes,
revealing  rapid oscillations in the case of smaller chains,
with the amplitude of the oscillations being close to its maximum value $S_{L/2}=L/2$ in the case of $L=4$ size chain  \cite{sunil1,sunil2}.
For large chain sizes the oscillation rate as well as the amplitude decline.
Comparing the two-tangles and the von~Neumann entropy for the four-spin system,
we see that these quantities are complementary to each other.
At $t=0$, the two-tangle and the von~Neumann entropy are maximal.
During the time evolution (after the kick), however,
the von~Neumann entropy attains the maximum value when the two-tangle is minimal and vice versa,
for all considered instances.
Thus the decrement of two qubit entanglement can be traced back to the rise of many party  entanglement sharing.
For larger chain sizes when the two-tangles vanish we expect the entanglement to exist in multiparticle form.
This may be the reason for the large von~Neumann entropy in the case of larger size chains.

\begin{figure}[t!]
 \includegraphics[width=0.40\textwidth]{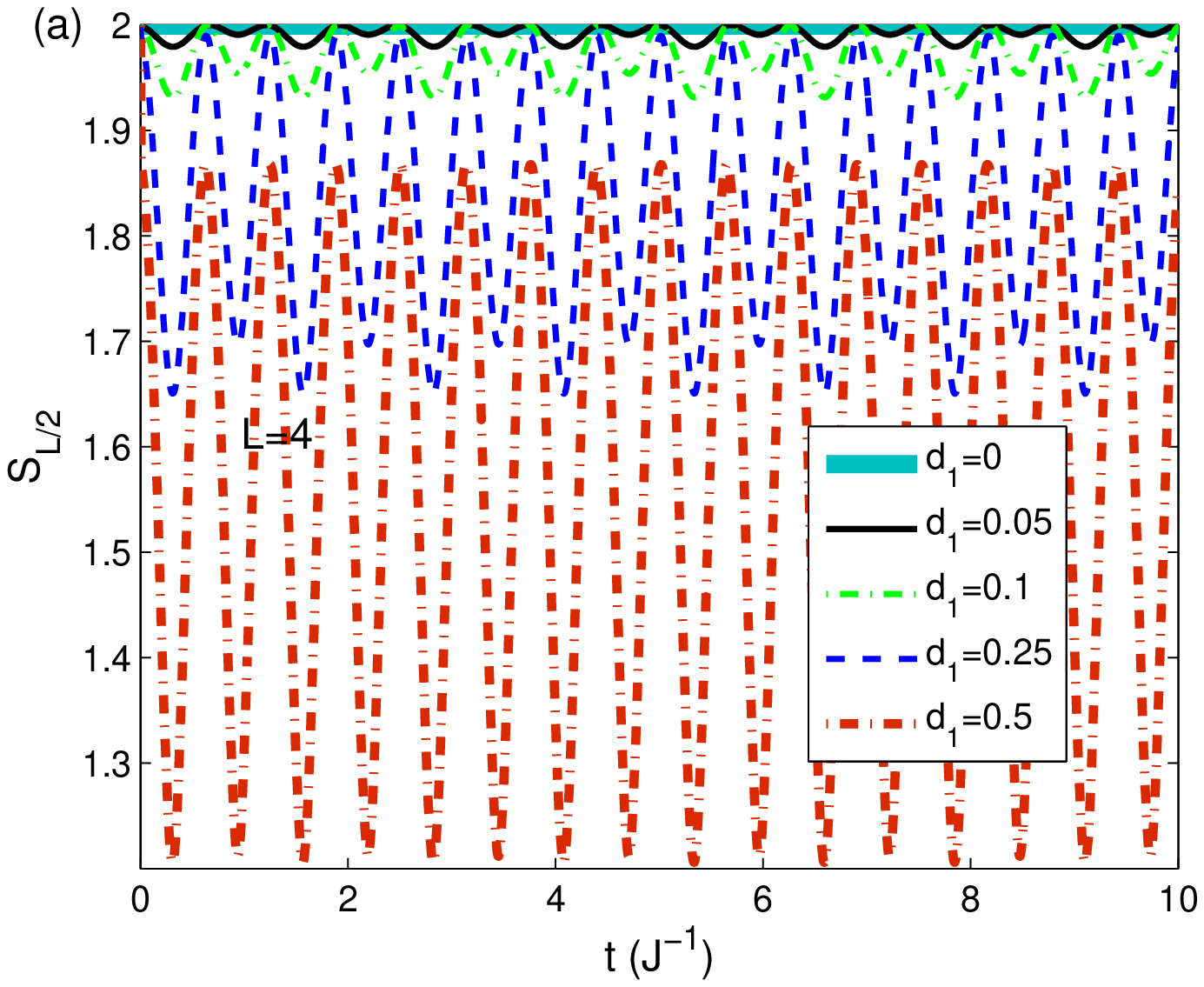}\\
 \includegraphics[width=0.40\textwidth]{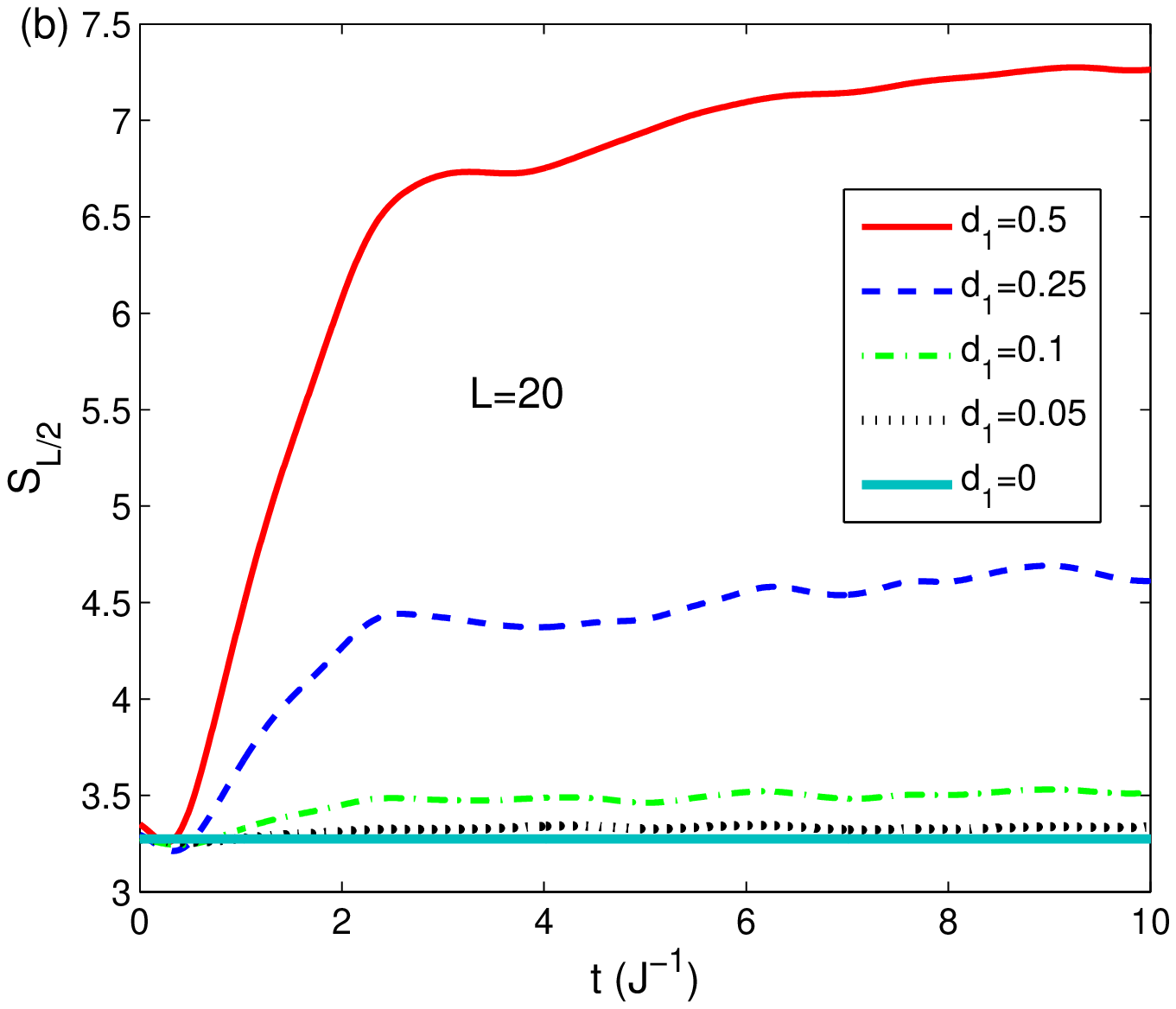}
 \caption{The time dependence of the von~Neumann entropy for (a) $L=4$ and (b) $L=20$ for different pulse strengths $d_1$.
          In all the cases the initial state is a state with $L/2$ number of excitations (i.e., $S^z=0$ sector).
          The duration of the applied pulse is $\varepsilon \approx 0.3$ and the strength $d_1=0.5$.
          The spin-exchange couplings and initial electric-field strength are $J_1=-J_2=-1.0$ and $d_0=0.05$, respectively.
          }
 \label{entro_l4L8}
\end{figure}
In Fig.~\ref{entro_l4L8} we compare chains with  $L=4$ and $L=20$,
analyzing the von~Neumann entropy as the pulse strength $d_1$ varies.
We choose the parameters such that the initial state is $L/2$-excitations spin state ($S^z=0$).
For $L=4$ (see Fig.~\ref{entro_l4L8}(a)) the initial state is two-excitation ground state.
In the absence of the pulse,
the von~Neumann entropy for $L=4$,
having the maximum at $t=0$, remains constant throughout the observation.
As we apply the pulse,
the entropy oscillates in time,
and the amplitude of oscillation grows proportionally to the pulse strength.
The maximal value, however, is always less than $S_{L/2}=2$.
For $L=20$ (see Fig.~\ref{entro_l4L8}(b)),
in the  absence of the pulse, the von~Neumann entropy -- as expected -- is again constant in time.
The value, however, is much smaller than the maximal possible entropy $S_{L/2}=L/2$.
While increasing the pulse strength, the entropy increases and saturates to a high value at a later time.
We also clearly observe the pattern of the linear growth of the entanglement entropy,
similar to the expected behavior for  global quenches \cite{Calabrese2005,Eisert2010}.
Hence, depending on the initial state, after the application of the pulse,
the system undergoes a transition to a superposition of  eigenstates of the Hamiltonian.
If the characteristic frequencies of the superposition state are non-commensurate, a partial loss of pair concurrence occurs.
For  larger systems this effect is more prominent. { The multiparticle entanglement  shows a robust behavior and survives if the characteristic frequencies of the superposition states are
incommensurate. This statement is valid for large systems as well.   }
We are mostly interested in the behavior of the chirality (see Fig.~\ref{chiral_L20}) which serves as an order parameter for our system. We observe that the chirality shows an oscillatory  behavior but the amplitude of the chirality slightly decays in the non-commensurate phase.

\section{\label{sec:DynamicalPT} Pulse and quench induced dynamical transition}
Connection between the canonical partition function $Z={\rm Tr}e^{-\beta H}$
and the return probability of a system to the initial state,
while going through a non trivial time evolution,
was  discussed in Ref.~[\onlinecite{heyl}].
This return probability
is also known as Loschmidt echo.
Nonanalyticity in time signifies a {\it dynamical phase transition} \cite{heyl}.
Here we consider the case of a sudden quench of the electric field.
The system is initially prepared at $t=0$  in the ground state of  $\hat{H}=\hat{H_0}+\hat{H_1}$,
and then it is suddenly quenched to $\hat{H}\to \hat{H}_0$ ($\hat{H_1}$ is also absent for $t>0$).
The quantity of interest is the Loschmidt echo $G(t)$  given as
 \begin{eqnarray}\label{gt_val}
  G(t)=\langle \psi_0|e^{-i\hat{H_0}t}|\psi_0\rangle,
 \end{eqnarray}
where $|\psi_0\rangle$ is the ground state of Hamiltonian (\ref{Hamiltonian2}).
The quantity $G(t)$  stands for the probability of returning to the ground state before the quench.

\begin{figure}[!t]
 \includegraphics*[width=0.4\textwidth]{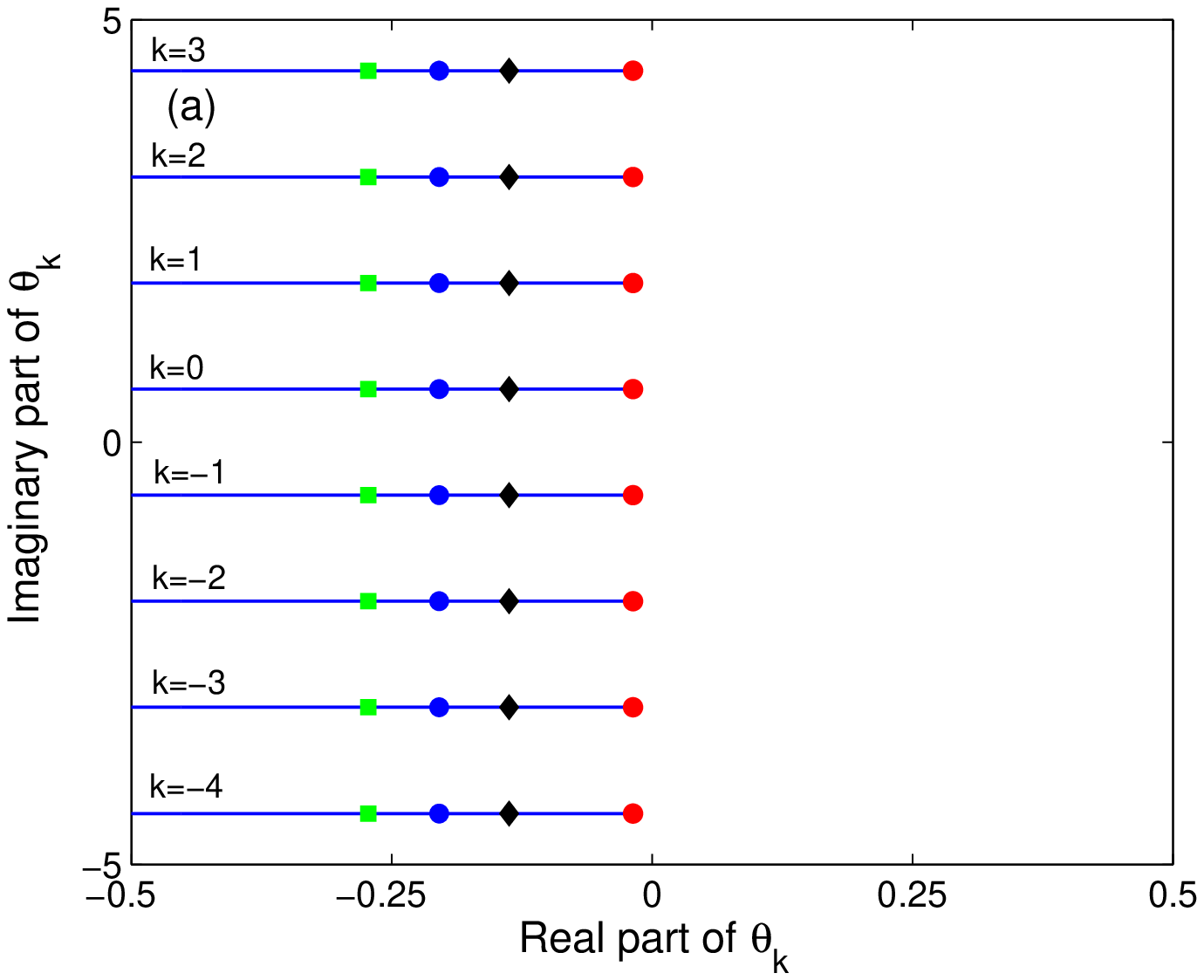}\\
 \includegraphics*[width=0.4\textwidth]{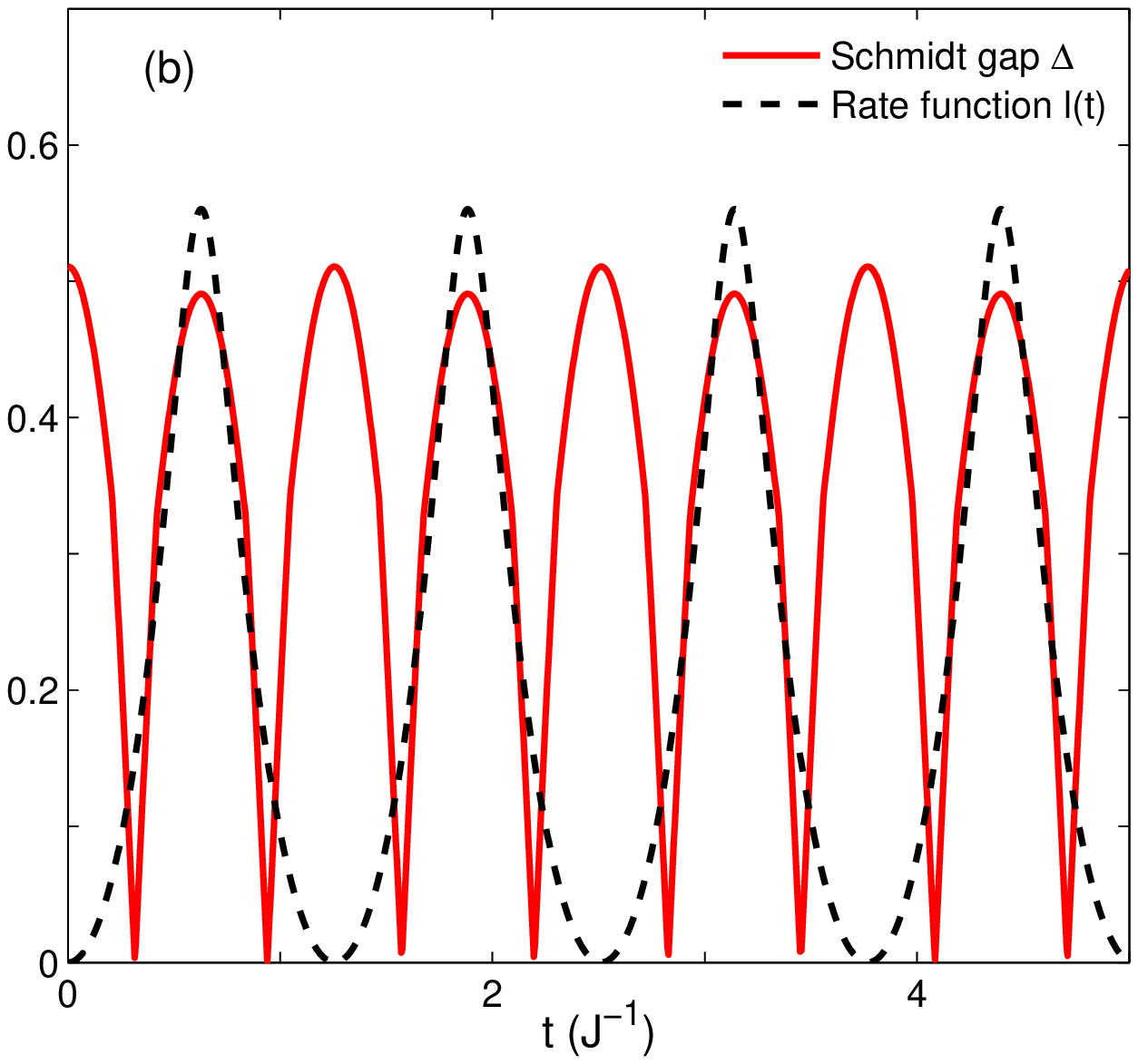}\\
 \includegraphics*[width=0.4\textwidth]{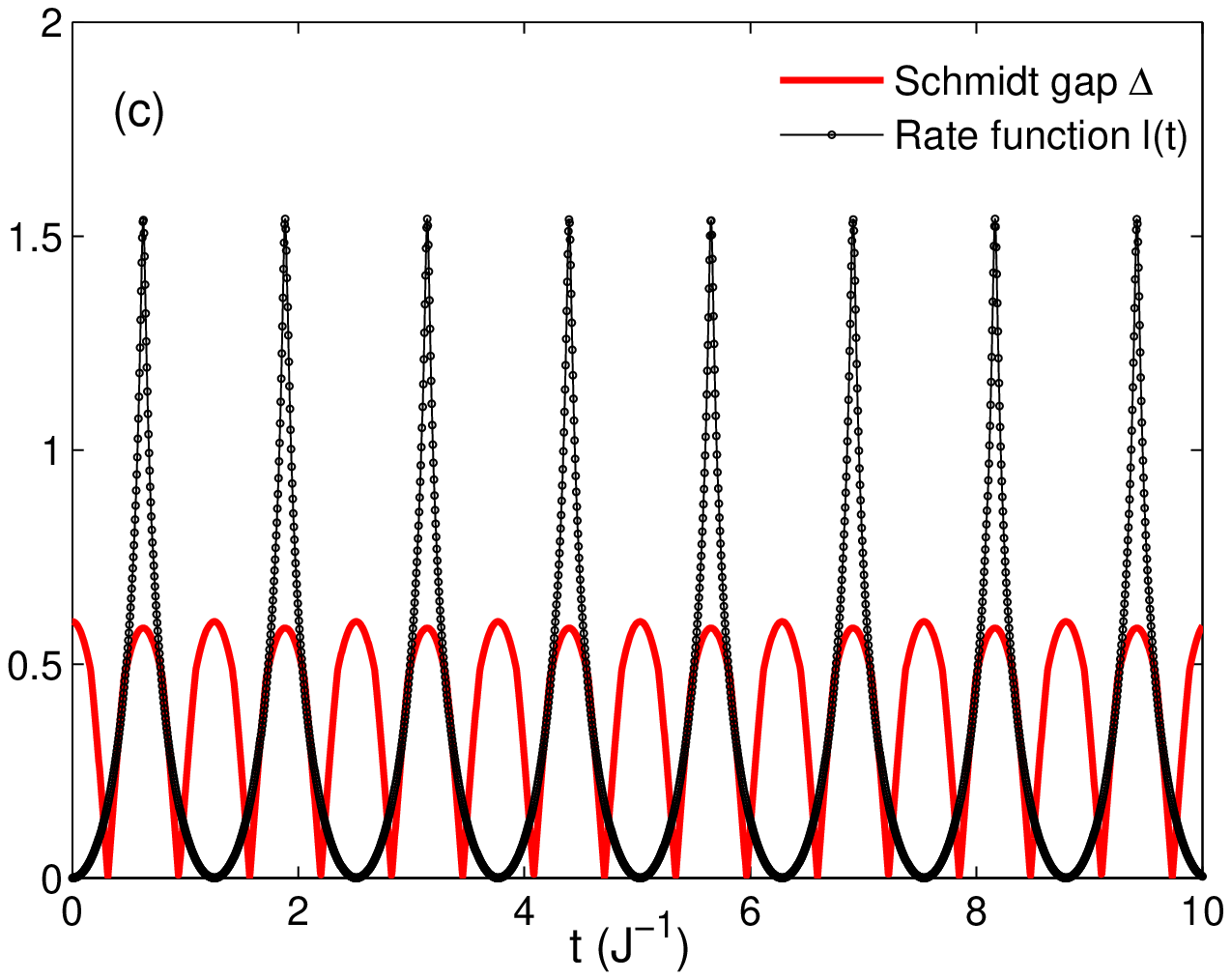}
 \caption{Quench protocol applied to 4-spin system in two-excitation ground state.
          (a) Zeros $\theta_k=it_k$ for different $k$
          and $J_1=-J_2=-1.0$, $B=0.25$, $d_0=0.05$.
          Green squares, blue circles, black diamonds, and red points
          correspond to $d_1=2.5$, $d_1=3.5$, $d_1=5.49$, and $d_1=100.0$ respectively.
          (b) Rate function and Schmidt gap for the parameters corresponding to black diamonds in (a).
          The Schmidt gap acquires the maximum at times $t$ where the rate function becomes zero.
          (c) Rate function and Schmidt gap for the parameters corresponding to red points in (a).
          }
 \label{quench_4_fig}
\end{figure}

\begin{figure}[!t]
 \includegraphics*[width=0.4\textwidth]{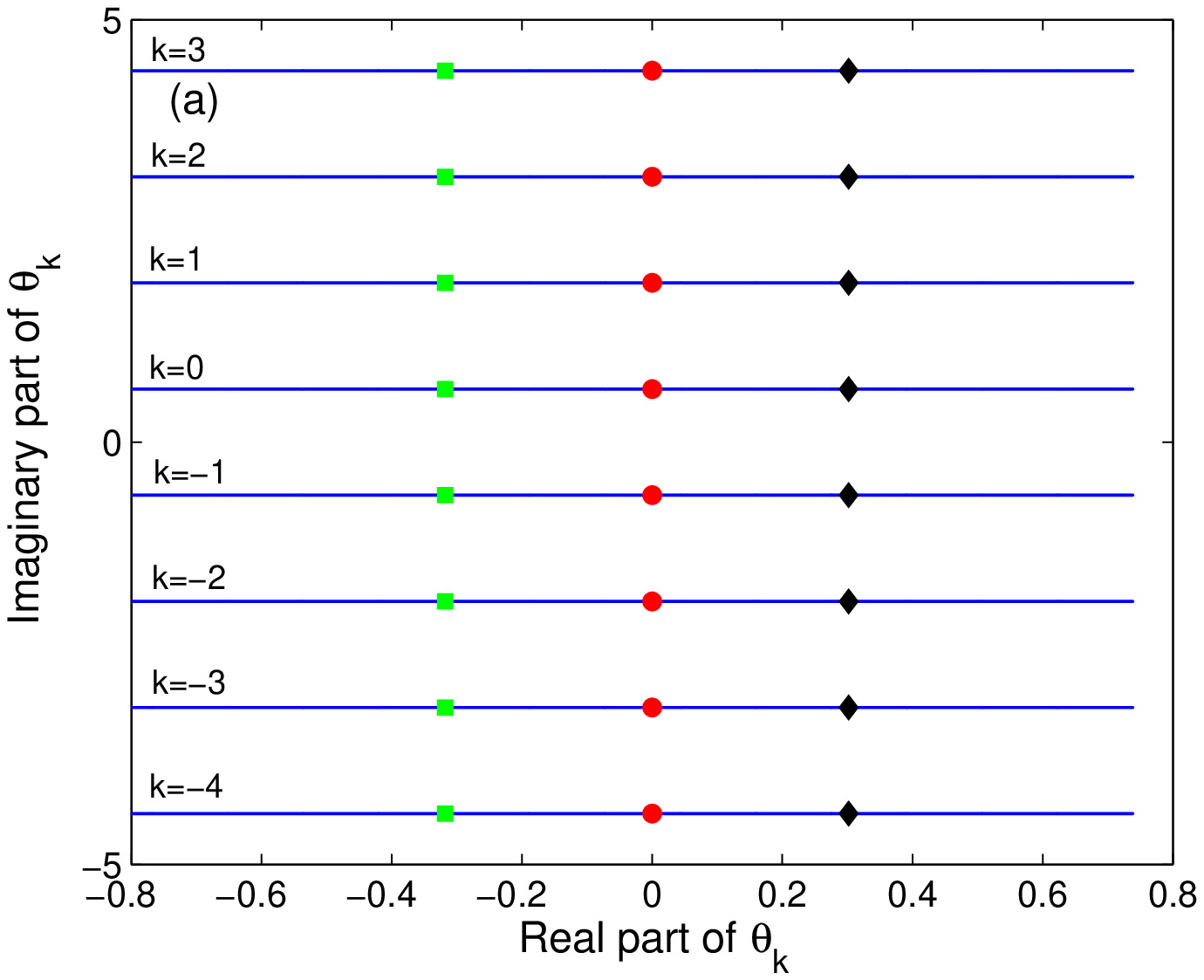}\\
 \includegraphics*[width=0.4\textwidth]{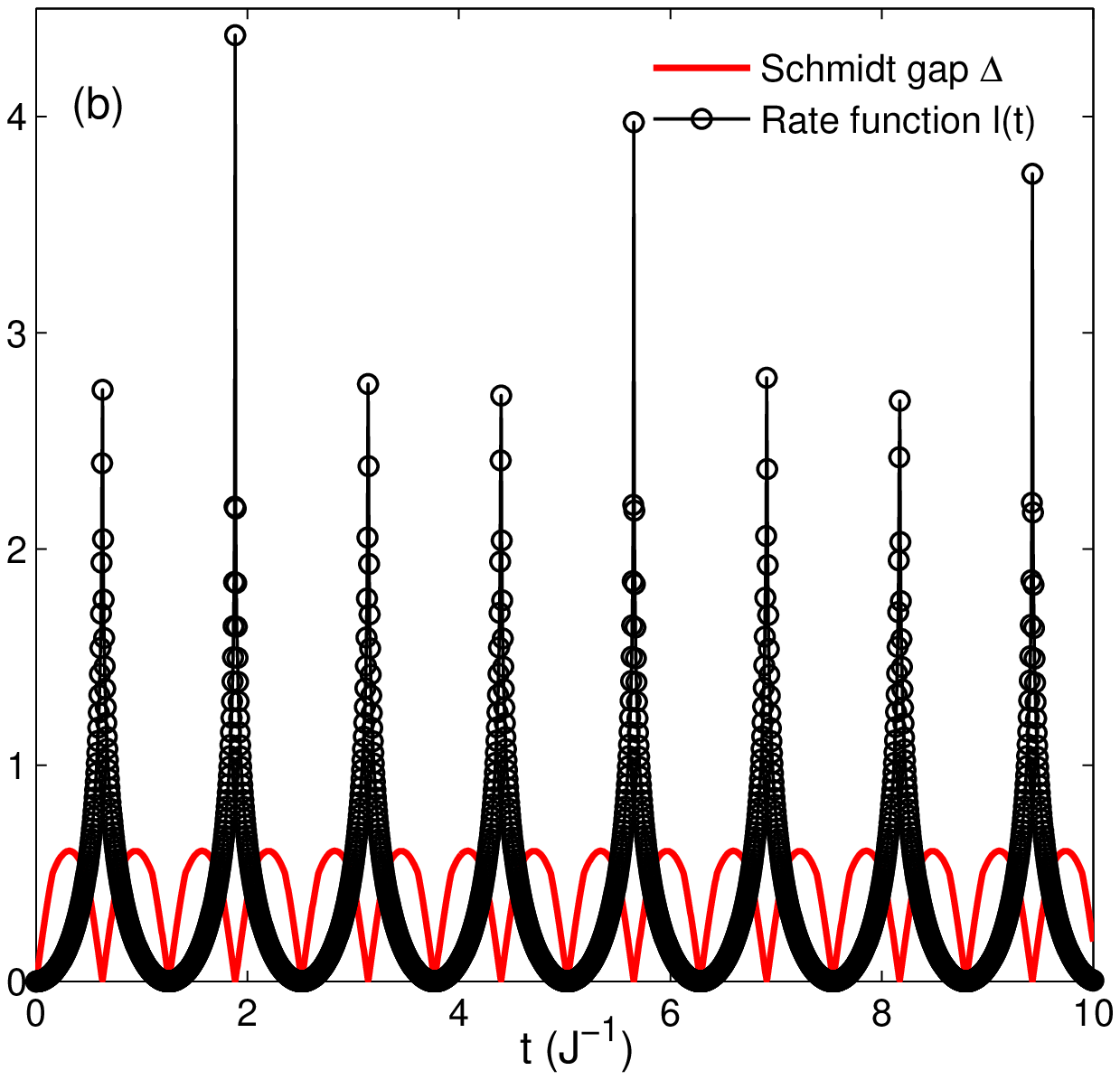}
 \caption{Pulse applied to 4-spin system in two-excitation ground state.
          (a) Zeros $\theta_k=it_k$ for different $k$ and $J_1=-J_2=-1.0$, $B=0.25$, $d_0=0.001$.
          Green squares, red points, and black diamonds
          correspond to $d_1=0.3$, $d_1=0.5554$, and $d_1=0.8$ respectively.
          In case of $d_1=0.5554$ (red points) $\Re(\theta_k)\approx 0$.
          (b) Rate function and Schmidt gap for $d_1=0.5554$.
          }
 \label{pulse_4_fig}
\end{figure}

\begin{figure}[!t]
 \includegraphics*[width=0.4\textwidth]{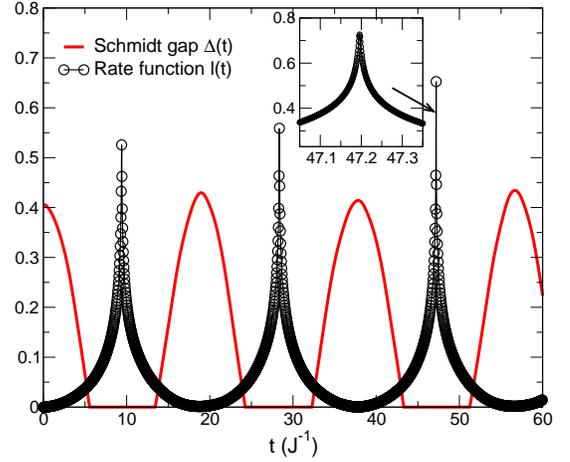}
 \caption{Time evolution of the rate function $l(t)$ and Schmidt gap $\Delta(t)$ for a periodic chain of size $L=22$.
          The peaks of the rate function at time $t^*\approx 9.41,\ 28.36,\ 47.20, \cdots$ correspond to the dynamical phase transitions.
          The inset shows a zoom into the cusp region at one of the nonanalytic point.
          The parameters are $J_1=-J_2=-1.0$.
          The electric fields $d_1$ and $d_0$ here $d_1=0.44, d_0=0.057$.}
 \label{quench_22_fig}
\end{figure}
We recall that the thermodynamic phase transitions in finite systems are indicated by changes in the zeros of the partition function $Z$
in the complex inverse temperature $\beta(=1/k_BT)$ plane (approaching the real $\beta$ axis in the thermodynamic limit).
The zeros of the Loschmidt echo $G(t)$ lie on real time axis.
Recent papers (see Refs.~[\onlinecite{karrasch,vajna,hickey,canovi}])
use the protocol by Heyl \emph{et. al.} \cite{heyl} to investigate dynamical phase transitions in various  models.
The quantity which is analogous to the thermodynamic free energy density is the rate function of the return probability given by
\begin{eqnarray}
 l(t) = -\lim_{L \to \infty}\frac{1}{L}\ln |G(t)|^2.
\end{eqnarray}

The singular points of $l(t)$ can be obtained by finding the zeros of $G(t)$.
In the thermodynamic limit, just as the free energy density manifests singularity rooted in the phase transition,
the rate function $l(t)$ reflects the singularity associated with dynamical phase transition.
Heyl\emph{ et al. } used this idea to study quenches in an transverse field Ising spin model,
which exhibits quantum phase transition between ferromagnetic and paramagnetic ground states.
Here we are interested in chiral MF spin chain close to magnetic saturation.
For the considered parameter regime $J_2=-J_1=1$,
the studied system might exist in the ferromagnetic, nematic, or chiral (VC) phase \cite{Azimi2}.

As it was shown above, for $L=4$ (which is analytically solvable),
one can observe effects qualitatively similar to a dynamical phase transition.
Namely singularity in the the rate function $l(t)$.
{We note, for small systems the singularity in the rate function is not rigorous criteria of phase transition}.
Only if the singularity survives in larger systems it is a signature of the phase transition.

This result we confirmed by exact numerical calculations for a system of $L=22$ spins.
The expression for  the Loschmidt echo
can be decomposed in the following form
\begin{eqnarray}\label{gteq}
 G(t)=\sum_{n,m}Q_{0n}H_{nm}Q_{m0},
\end{eqnarray}
where $Q_{n0}=\langle \phi_n|\psi_0\rangle$, $H_{nm}=\langle \phi_n|\exp(-i\hat{H}_0t)|\phi_m\rangle$,
and $|\phi_n\rangle$ is the $n^{\rm th}$ eigenstate of the Hamiltonian $\hat{H}_0$.
The zeros of the Weierstrass factorization are related to a dynamical phase transition \cite{heyl}.
According to Weierstrass theorem for entire functions, an entire function $f(z)$ with the zeros $z_j$, $j=1,2,3,\cdots$
can be written as
\begin{eqnarray}
 f(z)=e^{g(z)}\prod_j\left(1-\frac{z}{z_j}\right).
\end{eqnarray}
Here $g(z)$ is another entire function of $z$.
In the above equation we see that the singular (nonanalytic) part of the function $\ln|f(z)|$ is exclusively determined by zeros $z_j$.
Considering $z=it$ and $f(z)$ as $G(t)$ (given by Eq.~\eqref{gteq}),
the rate function becomes
\begin{eqnarray}
 l(t)=-\frac{2}{L}\left[|g(t)|+\sum_j\ln\left|1-\frac{t}{t_j}\right|\right].
\end{eqnarray}
In the case of four-spin chain with two-excitations,
the ground state is $\vert \phi_7\rangle$ and the return probability is given by
\begin{eqnarray}\label{two_excitation_State_quench}
 G(t)=ae^{-i\mathcal{E}_6t} +be^{-i\mathcal{E}_7t}.
\end{eqnarray}
The eigenstate $\vert \phi_7\rangle$ and full details on the used notations $a=[\alpha\gamma^{\prime}(4+2\eta\lambda^{\prime})]^2$,
$b=[\gamma\gamma^{\prime}(4+2\lambda\lambda^{\prime})]^2$ are given in the Appendix~\ref{sec:appA}.
The parameters  $\lambda^{\prime}$ and $\gamma^{\prime}$ are obtained via substituting $d_0$ with $d_1+d_0$ in Eq.~\eqref{eq:A2}.
$G(t)=0$ at
\begin{eqnarray}
 t_k=\frac{1}{\mathcal{E}_7-\mathcal{E}_6}\left(i\ln\left(\frac{a}{b}\right)-\pi(2k+1)\right),
\end{eqnarray}
where $k=0,\pm1, \pm2,\cdots$.
Real and imaginary parts of $\theta_k=it_k$ can be written separately as
\begin{eqnarray}
 \Re(\theta_k)=\frac{2}{\sqrt{(J_1-4J_2)^2+8d_0^2}}\ln\left(\left|\frac{\alpha(2+\eta\lambda^{\prime})}{\gamma(2+\lambda\lambda^{\prime})}\right|\right)\nonumber \\
\end{eqnarray}
and
\begin{eqnarray}\label{Imaginary}
 \Im(\theta_k)=\frac{\pi(2k+1)}{\sqrt{(J_1-4J_2)^2+8d_0^2}}.
\end{eqnarray}
The obtained analytical results, which are plotted in Fig.~\ref{quench_4_fig},
indicate an onset of  a dynamical transition.
However, small ($L=4$) spin systems do  not possess  a well-developed phase-transition behavior,
but precursors that hint on a dynamical transition. Signature of quantum phase
transition can be observed in the case of a pulse induced dynamics as well.
Using $\vert\phi_7\rangle$ as an initial ground state (cf. with Eq.~\eqref{two_excitation_State_quench} for the quench scenario)
and (\ref{eq:B4}), (\ref{eq:B5}) (see Appendix~\ref{sec:appB}) we plot the rate function and the Fisher zeros on Fig.~\ref{pulse_4_fig}.
As we see the nonanalytic, i.e. the singular or cusp-like behavior of the rate function is correlated with the minimum of Schmidt gap and the Fisher zeros cross the real axis.
The Schmidt gap, defined as a difference between two largest eigenvalues of the reduced density matrix of the bipartite chain,
is one of the witnesses of the dynamical phase transition \cite{hickey}.

\begin{figure}[!t]
\includegraphics*[width=0.48\textwidth]{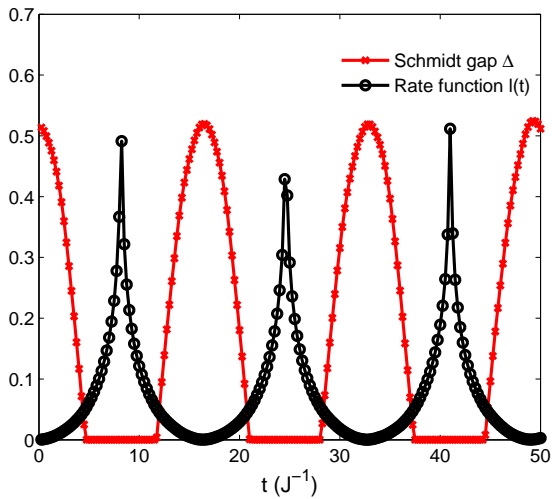}
\caption{The rate function and the Schmidt gap for a periodic chain with size $L=18$ and $J_1=-J_2=-1.0$.
         The pattern of the Schmidt gap and the rate function are complementary to each other.
        For $L=18$ the electric fields $d_1$ and $d_0$ are chosen  as $d_1=2.34, d_0=0.098$, respectively.}
\label{quench_18_fig}
\end{figure}

Let us inspect the dynamical phase transition by calculating the time evolution of the rate function, $l(t)$ for larger systems.
Fig.~\ref{quench_22_fig} shows results for a chain with $L=22$ spins and an initial state taken  to be the ground state in two-excitation sector (two spins flipped in magnetically saturated state). For the parameter set $\{d_1=0.44,\ d_0=0.057\}$,
we see cusps at times $t^*\approx 9.41$, $28.36$, $47.20,\ \cdots$,
where the rate function, $l(t)$, develops cusps, indicating singularities.
These points may refer to a transition between different type of chiral order signifying  a dynamical phase transition.
Yet another pairs $\{d_1,\,d_0\}$ can also be found for which the similar transitions take place.
For example on Fig.~\ref{quench_18_fig} we show results for $L=18$ size chain with $d_1=2.34$ and $d_0=0.098$.
Another quantity, that might be employed for the detection of the dynamical phase transitions is the so called Schmidt gap,
which is related to the entanglement spectrum \cite{hickey}.
In 1D, the ground-state entanglement entropy shows a logarithmic behavior with the system size \cite{Eisert2010}.
At the quantum-critical point $S\sim c\log l$ and diverges logarithmically with the block size $\ell$.
Here $c$ is the central charge of the corresponding conformal field theory describing the quantum phase transition \cite{Eisert2010}.
Close to the quantum-critical point $S\sim c\log \xi$ and $\xi$ is the correlation length.
After the quench, however, the system, in general, will be in  the excited state not the ground one.
Entanglement entropy in the excited state shows a qualitatively different behavior.
Namely, after the global quench the entanglement entropy  grows typically linearly in time \cite{Calabrese2005,Eisert2010,Sirker3}.
Away from the critical point the system can be characterized by the entanglement spectrum \cite{Li_haldane,chiara1,chiara2,chandran}
i.e., the eigenvalues of the reduced density matrix of one of the two partitions (Schmidt eigenvalues)
while tracing out the degrees of freedom of the other part.
The entanglement spectrum is an accepted tool to characterize the many body system.
However, the information about the quantum critical points in a many body system,
can be easily extracted by merely knowing the gap between the two largest eigenvalues,
the so called Schmidt gap, and without the knowledge of the full  entanglement spectrum.
The zeros of the Schmidt gap  provide information on the quantum critical point \cite{chiara1,hickey}.
We studied the time evolution of the Schmidt gap $\Delta=\lambda_1-\lambda_2$ ($\lambda_1$ and $\lambda_2$ are the two largest eigenvalues
of the reduced density matrix) for all cases followed in this section.
Fig.~\ref{quench_22_fig} and Fig.~\ref{quench_18_fig} (as well as Fig.~\ref{quench_4_fig} and Fig.~\ref{pulse_4_fig})
also show the Schmidt gap, in addition to the rate function $l(t)$.
For the parameters, $d_1$ and $d_0$, for which there is a dynamical phase transition,
nonanalyticity at the points $t^*$ in the rate function,
we see a nice pattern in the Schmidt gap too.
In Fig.~\ref{quench_22_fig} we can extract the role of the Schmidt gap at the critical points.
We see that at the onset of the dynamical phase transition,
the Schmidt gap vanishes and remains zero in the time interval in which
the system undergoes a dynamical phase transition.
We can estimate the critical point by just looking at the pattern of the Schmidt gap.
For example in Fig.~\ref{quench_22_fig}, the Schmidt gap is zero for $5.56<t<13.42$
and the critical point $t^*=9.41$ is close to the middle $t=9.49$ of the interval.
At times when the rate function touches the minimum, the Schmidt gap reaches its maximum.
For $L=4$  the Schmidt gap only closes at the dynamical-transition points (singularities  of $l(t)$),
for quench (see Fig.~\ref{quench_4_fig}(c)) as well as the pulse (see Fig.~\ref{pulse_4_fig}(b)) scenarios.

\begin{figure}[!t]
 \includegraphics*[width=.45\textwidth]{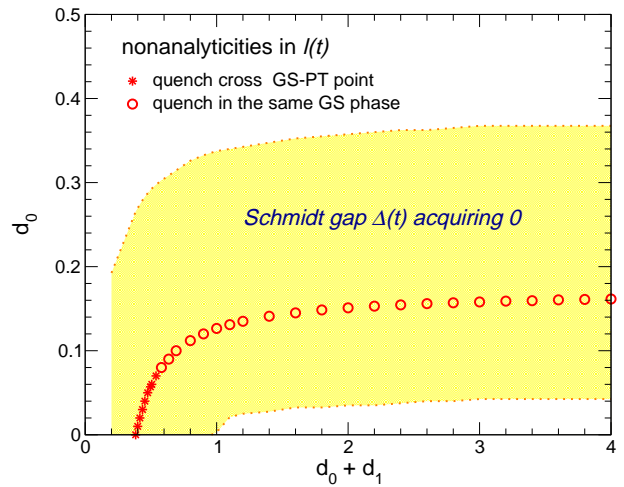}
 \caption{The diagram of the quench-parameter pairs $(d_0 + d_1)\rightarrow d_0$,
          for $L=22$ site system in the two-excitation sector.
          The parameter region (yellow area), for which the Schmidt gap acquires zeros in time,
          defines pairs of quench-parameters for which the dynamical phase transition occurs.
          We also show the quench-parameter pairs for which the rate function $l(t)$ has singularities
          (red asterisks and circles).
          The asterisks correspond to the quench between the different phases while the circles represent quench within the same phase.
          Further system parameters are $J_1=-J_2=-1.0$.
          }
 \label{quench_d_22}
\end{figure}

\begin{figure}[t]
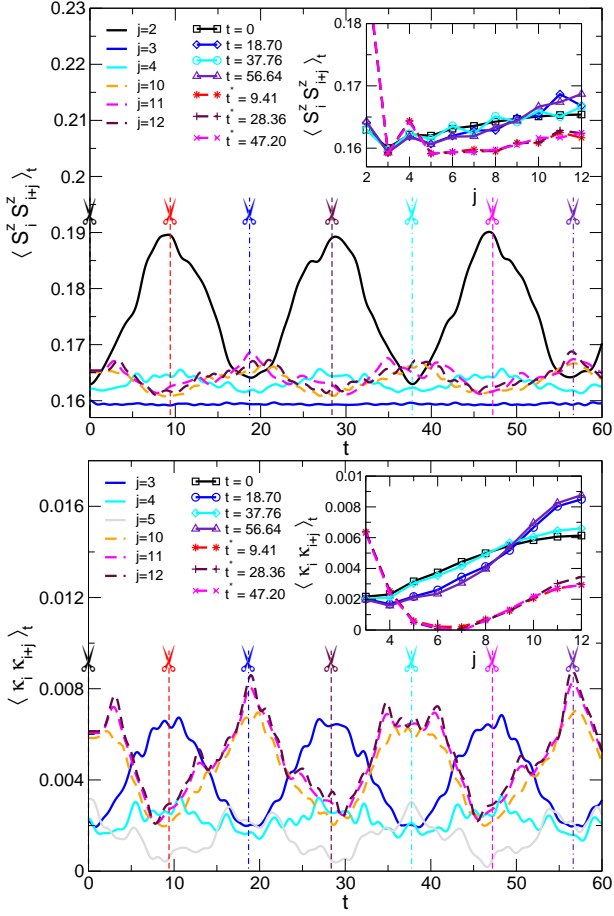

 \includegraphics*[width=0.45\textwidth]{szsz_t_22.eps}\\
 \includegraphics*[width=0.45\textwidth]{chch_t_22.eps}
 \caption{The time evolution of the equal-time spin-spin $\langle S^z_i S^z_{i+j} \rangle_t$ (upper panel)
          and the chirality $\langle \kappa_i \kappa_{i+j} \rangle_t = \langle(\vec{S}_{i}\times\vec{S}_{i+1})^z(\vec{S}_{i+j}\times\vec{S}_{i+j+1})^z\rangle_t$
          (lower panel) correlation functions,
          for the case shown on Fig.~\ref{quench_22_fig}.
          Insets show the correlation functions vs.~distance $j$ (profiles of the cuts)
          at the singularities $t^*$  (indicated by dashed lines)
          and at local minima (indicated by dash-doted lines) of the rate function $l(t)$.
         }
 \label{corr_t_fig}
\end{figure}

\begin{figure}[t]
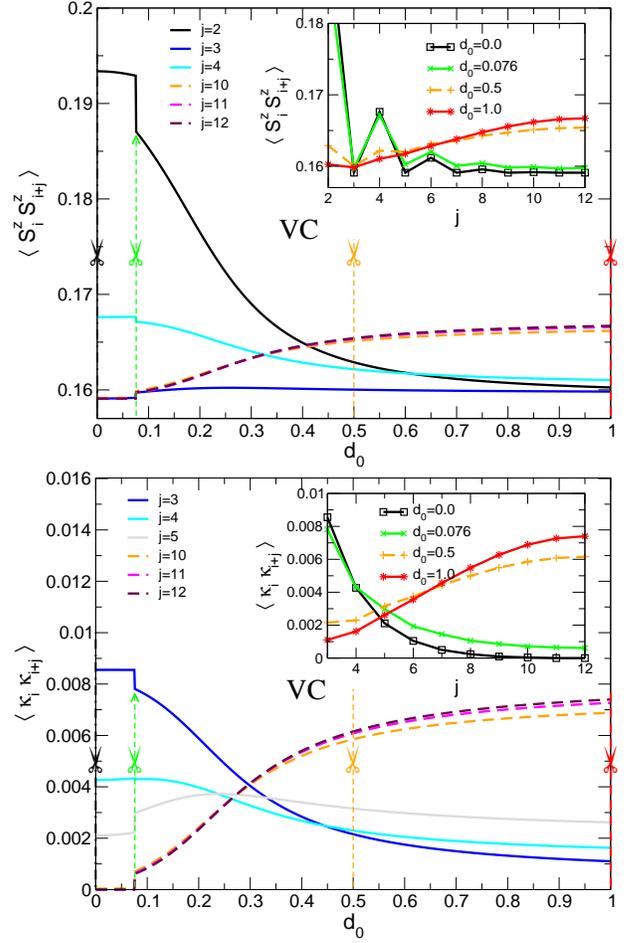

 \includegraphics*[width=0.45\textwidth]{szsz_22.eps}\\
 \includegraphics*[width=0.45\textwidth]{chch_22.eps}
 \caption{The spin-spin $\langle S^z_i S^z_{i+j} \rangle$ (upper panel)
          and  the chirality $\langle \kappa_i \kappa_{i+j} \rangle = \langle(\vec{S}_{i}\times\vec{S}_{i+1})^z(\vec{S}_{i+j}\times\vec{S}_{i+j+1})^z\rangle$
          (lower panel) the correlation functions vs. $d_0$,
          for $L=22$ spin system in two-excitation ground state, $J_1=-1.0$.
          Insets show the correlation functions vs.~distance $j$ (profiles of the cuts)
          at $d_0=0.0$, $0.076$, $0.5$, $1.0$ indicated by dashed lines in the main plot.
          One can also identify the level-crossing transition exhibited in the finite jump in the correlation functions
          at the transition point $d_0\approx 0.076$.
         }
 \label{corr_fig}
\end{figure}

Analyzing the behavior of the Schmidt gap ($\Delta(t)$), as well as singularities of the rate function ($l(t)$),
after the quench $(d_0+d_1 \rightarrow d_0)$,
we constructed the diagram of the quench-parameter pairs, shown on Fig.~\ref{quench_d_22}.
The region of the quench parameters, for which the Schmidt gap acquires zeros during the time-evolution of the system,
identifies the quench-parameter sets for which the dynamical phase transition occurs.
This region in the parameter space is quite large and includes dynamical transitions,
detected by singularities in the rate function.

In order to gain more insight into the time evolution of the system,
in Fig.~\ref{corr_t_fig} we show the equal-time spin-spin $\langle S^z_i S^z_{i+j} \rangle_t$
and chirality $\langle \kappa_i \kappa_{i+j} \rangle_t = \langle(\vec{S}_{i}\times\vec{S}_{i+1})^z(\vec{S}_{i+j}\times\vec{S}_{i+j+1})^z\rangle_t$
correlation functions, for the three closest and the three furthest sites.
The shown data corresponds to the above considered case of $L=22$ spin chain, close to magnetic saturation (two-excitations sector).
In the chirality correlator, one clearly identifies the emergence of two distinct behavior.
We also show the entire profile of the chirality correlator at selected sets of time points,
as an inset of the corresponding plot,
that completely underlines this scenario.
While long-range chirality correlation functions are suppressed for the time points $t^*$ at which the rate function ($l(t)$) has singularities,
the short range chirality correlations build-up at the same time points.
The long-range chirality order is recovered at time points where the rate function is zero.
The data shown on Fig.~~\ref{corr_t_fig}, also reveals the appearance of two distinct long-range behavior.
The true recovery of the initial phase is only happening at each second zero of the rate function.
The alteration between short- and long-range ordered phases,
is supported by the behavior of the same correlator functions in the ground state of the system at different electric-field strengths.
On Fig.~\ref{corr_fig} we show the spin-spin and the chirality correlators as a function of the electric-coupling strength $d_0$,
for the three closest and the three furthermost sites, as well as the entire correlator profiles for the selected electric-coupling strength points (see the insets on Fig.~\ref{corr_fig}).
The ground-state correlation functions clearly reveal the level crossing phase transition
in the system at $d_0\approx0.076$ upon increasing $d_0$.
This transition corresponds to the static phase transition from the nematic, two-magnon bound state to the long-range vector chiral (VC) state \cite{Azimi2,Chubukov2001}.
While for $d_0 < 0.076$ both the spin-spin as well as the chirality correlation functions are short-ranged,
the system develops a long-range chirality order after the transition to the VC phase, for $d_0 > 0.076$.
Therefore, we conclude, that a system  prepared initially  in the long-range chirality ordered phase (e.g., as for $d_1 + d_0 = 0.497$),
undergoes after the quench (to $d_0=0.057$)  a dynamical phase transition to the nematic phase.
The corresponding phase is characterized by the absence of a long-range chiral order (cf. Fig~\ref{corr_t_fig}(b) with Fig~\ref{corr_fig}(b)).
Note that a similar dynamical phase transition can also occur
when the system is quenched within the same static phase \cite{Sharma2015},
namely, in the long-range VC phase,
as it is shown on Fig.~\ref{quench_d_22} (e.g., the red circles).
The level crossing transition persists in all systems with $L>4$ spins in the case of the two-excitation ground state.

The ferroelectric polarization of the MF system couples to a spatially uniform electrical field as
$-\vec{E}(t)\cdot\vec{P}=E(t)g_{\mathrm{ME}} \sum_{i=1}^L(\vec{S}_{i}\times\vec{S}_{i+1})^z$ (cf. Eq.~\eqref{Hamiltonian1}) which vanishes in the absence of a long-range chiral order.
The alteration between the short- and long-range chirality order
 at the verge of  the dynamical phase transition
might  be detected experimentally by monitoring the polarization dynamics of the MF system (via the associated emission spectra).

\section{\label{sec:Conclusions}Conclusions}
We studied a multiferroic spin chain  subjected to   short electric-field pulses in addition to static electric and magnetic fields.
The electric field pulse  switches the system, initially in the ground state, into incommensurate chiral phase.
(superposition of the excited states with incommensurate spectral properties $\omega_{i}=E_{i}-E_{0}/\hbar,~~\omega_{i}/\omega_{j}\neq \mathrm{integer}$). The time evolution of chirality  signifies the  emergence of a spin configuration in new phase.
Preparing the initial state of the system in $n$-excitation spin state, by manipulating electric and magnetic fields,
we analytically studied a model of four spins and calculated the chirality,
the one-tangle, the two-tangle and the von~Neumann entropy.
We find  that  all the measured quantities are constants for the one-excitation ground state
and evolve in time in $n\left(>1\right)$-excitation ground state.
Employing exact diagonalization methods for systems with $L>4$ spins,
we found that the two-tangle  vanishes as the system size increases.
The decay of the two-tangle is related to the non-commensurate characteristic frequencies of the superposition state.
For larger systems this effect is more prominent.
The chirality oscillates with time, but its peak value is not vanishing as compared to the case of the two-tangle which vanishes for systems with $L>4$ spins.
The linear growth of the von~Neumann entropy is also observed for  systems with $L>4$ spins after the pulse is applied,
confirming the fact that the system is in a superposition of excited states.
The computed data shows that the many particle entanglement and the chirality are robust
and persist in the incommensurate phase even for larger systems.
We also employed a quench protocol in order to calculate the Loschmidt echo
and the rate function to inspect the dynamical phase transitions between the chiral and nematic phases.
Quenching the system suddenly,
we calculate the return probability to the pre-quench ground state.
Signatures of  dynamical phase transitions are identified and observed.
The critical points are those where the rate function (which is analogous to thermodynamic free energy density) is nonanalytic.
The zeros of the Loschmidt echo resembles the zeros of the partition function.
The Schmidt gap is also sensitive to these quantum critical points
where the system approaches nematic phase from the long-range VC phase.
We clearly observed alteration between short- and long-range chirality phases that occurs at the onset of
dynamical phase transitions.

\section*{Acknowledgements}
We would like to thank M. Heyl and L. Chioncel for useful discussions and valuable comments.
Financial support by the Deutsche Forschungsgemeinschaft (DFG) through SFB 762, is gratefully acknowledged.
MS acknowledges support by the Rustaveli national science foundation through the grand no. FR/265/6-100/14.
SKM acknowledges Department of Science and Technology,
India for support under the grant of INSPIRE Faculty Fellowship.
\newpage
\appendix
\section{\label{sec:appA}Eigenfunctions and eigenvalues}
Eigenfunctions and eigenvalues of the Hamiltonian
$\hat{H_0}$ (Eq.~\eqref{ham0}) with $d_0=E_0g^{\phantom{+}}_{\mathrm{ME}}$ in the case of four spins read
\begin{eqnarray}\label{eq:A1}
 \vert\phi_{1}\rangle&=&\vert\uparrow\uparrow\uparrow\uparrow\rangle,\nonumber\\
 \vert\phi_{2}\rangle&=&\frac{i}{2}\vert\downarrow\uparrow\uparrow\uparrow\rangle+\frac{-1}{2}\vert\uparrow\downarrow\uparrow\uparrow\rangle
 +\frac{-i}{2}\vert\uparrow\uparrow\downarrow\uparrow\rangle+\frac{1}{2}\vert\uparrow\uparrow\uparrow\downarrow\rangle,\nonumber\\
 \vert\phi_{3}\rangle&=&\frac{-i}{2}\vert\downarrow\uparrow\uparrow\uparrow\rangle+\frac{-1}{2}\vert\uparrow\downarrow\uparrow\uparrow\rangle
 +\frac{i}{2}\vert\uparrow\uparrow\downarrow\uparrow\rangle+\frac{1}{2}\vert\uparrow\uparrow\uparrow\downarrow\rangle,\nonumber\\
 \vert\phi_{4}\rangle&=&\frac{1}{2}\vert\downarrow\uparrow\uparrow\uparrow\rangle+\frac{-1}{2}\vert\uparrow\downarrow\uparrow\uparrow\rangle
 +\frac{1}{2}\vert\uparrow\uparrow\downarrow\uparrow\rangle+\frac{-1}{2}\vert\uparrow\uparrow\uparrow\downarrow\rangle,\nonumber\\
  \vert\phi_5\rangle&=&\frac{1}{2}\vert\downarrow\uparrow\uparrow\uparrow\rangle+\frac{1}{2}\vert\uparrow\downarrow\uparrow\uparrow\rangle
 +\frac{1}{2}\vert\uparrow\uparrow\downarrow\uparrow\rangle+\frac{1}{2}\vert\uparrow\uparrow\uparrow\downarrow\rangle,\nonumber\\
 \vert\phi_6\rangle&=&\alpha\big(\vert\downarrow\downarrow\uparrow\uparrow\rangle-i\eta\vert\downarrow\uparrow\downarrow\uparrow\rangle
 -\vert\downarrow\uparrow\uparrow\downarrow\rangle-\vert\uparrow\downarrow\downarrow\uparrow\rangle\nonumber\\
     &&+i\eta\vert\uparrow\downarrow\uparrow\downarrow\rangle+\vert\uparrow\uparrow\downarrow\downarrow\rangle\big),\nonumber\\
 \vert\phi_7\rangle&=&\gamma\big(\vert\downarrow\downarrow\uparrow\uparrow\rangle-i\lambda\vert\downarrow\uparrow\downarrow\uparrow\rangle
 -\vert\downarrow\uparrow\uparrow\downarrow\rangle-\vert\uparrow\downarrow\downarrow\uparrow\rangle\nonumber\\
      &&+i\lambda\vert\uparrow\downarrow\uparrow\downarrow\rangle+\vert\uparrow\uparrow\downarrow\downarrow\rangle\big),\nonumber\\
 \vert\phi_8\rangle&=&\frac{1}{\sqrt{6}}\big(\vert\downarrow\downarrow\uparrow\uparrow\rangle+\vert\downarrow\uparrow\downarrow\uparrow\rangle
 +\vert\downarrow\uparrow\uparrow\downarrow\rangle+\vert\uparrow\downarrow\downarrow\uparrow\rangle\nonumber\\
      &&+\vert\uparrow\downarrow\uparrow\downarrow\rangle+\vert\uparrow\uparrow\downarrow\downarrow\rangle\big),\nonumber\\
 \vert\phi_9\rangle&=&\frac{1}{\sqrt{12}}\big(\vert\downarrow\downarrow\uparrow\uparrow\rangle-2\vert\downarrow\uparrow\downarrow\uparrow\rangle
 +\vert\downarrow\uparrow\uparrow\downarrow\rangle+\vert\uparrow\downarrow\downarrow\uparrow\rangle\nonumber\\
      &&-2\vert\uparrow\downarrow\uparrow\downarrow\rangle+\vert\uparrow\uparrow\downarrow\downarrow\rangle\big),\\
 \vert\phi_{10}\rangle&=&\frac{-1}{\sqrt{2}}\vert\downarrow\downarrow\uparrow\uparrow\rangle
 +\frac{1}{\sqrt{2}}\vert\uparrow\uparrow\downarrow\downarrow\rangle, \nonumber\\
 \vert\phi_{11}\rangle&=&\frac{-1}{\sqrt{2}}\vert\downarrow\uparrow\uparrow\downarrow\rangle
 +\frac{1}{\sqrt{2}}\vert\uparrow\downarrow\downarrow\uparrow\rangle,\nonumber\\
 \vert\phi_{12}\rangle&=&\frac{i}{2}\vert\downarrow\downarrow\downarrow\uparrow\rangle+\frac{-1}{2}\vert\downarrow\downarrow\uparrow\downarrow\rangle
 +\frac{-i}{2}\vert\downarrow\uparrow\downarrow\downarrow\rangle+\frac{1}{2}\vert\uparrow\downarrow\downarrow\downarrow\rangle,\nonumber\\
 \vert\phi_{13}\rangle&=&\frac{-i}{2}\vert\downarrow\downarrow\downarrow\uparrow\rangle+\frac{-1}{2}\vert\downarrow\downarrow\uparrow\downarrow\rangle
 +\frac{i}{2}\vert\downarrow\uparrow\downarrow\downarrow\rangle+\frac{1}{2}\vert\uparrow\downarrow\downarrow\downarrow\rangle,\nonumber\\
 \vert\phi_{14}\rangle&=&\frac{1}{2}\vert\downarrow\downarrow\downarrow\uparrow\rangle+\frac{1}{2}\vert\downarrow\downarrow\uparrow\downarrow\rangle
 +\frac{1}{2}\vert\downarrow\uparrow\downarrow\downarrow\rangle+\frac{1}{2}\vert\uparrow\downarrow\downarrow\downarrow\rangle,\nonumber\\
 \vert\phi_{15}\rangle&=&\frac{1}{2}\vert\downarrow\downarrow\downarrow\uparrow\rangle+\frac{-1}{2}\vert\downarrow\downarrow\uparrow\downarrow\rangle
 +\frac{1}{2}\vert\downarrow\uparrow\downarrow\downarrow\rangle+\frac{-1}{2}\vert\uparrow\downarrow\downarrow\downarrow\rangle,\nonumber\\
 \vert\phi_{16}\rangle&=&\vert\downarrow\downarrow\downarrow\downarrow\rangle\nonumber,
\end{eqnarray}
\begin{eqnarray}
 &&\mathcal{E}_{1} =J_1+J_2-2B,\nonumber\\
 &&\mathcal{E}_{2} =-J_2-B-d_0, \qquad \mathcal{E}_{3}=-J_2-B+d_0,\nonumber\\
 &&\mathcal{E}_{4} =-J_1+J_2-B, \qquad \mathcal{E}_{5}=J_1+J_2-B,\nonumber\\
 &&\mathcal{E}_6   =-\frac{1}{2}\left({J_1}-2J_2-\sqrt{({J_1}-4{J_2})^2+8d_0^{2}}\right),\nonumber\\
 &&\mathcal{E}_7   =-\frac{1}{2}\left({J_1}-2J_2+\sqrt{({J_1}-4{J_2})^2+8d_0^{2}}\right),\nonumber\\
 &&\mathcal{E}_8   =J_1+J_2, \qquad \mathcal{E}_9=-2J_1+J_2,\nonumber\\
 &&\mathcal{E}_{10}=-J_2, \qquad \quad \mathcal{E}_{11}=-J_2,\nonumber\\
 &&\mathcal{E}_{12}=-J_2+B+d_0, \quad ~\mathcal{E}_{13}=-J_2+B-d_0,\nonumber\\
 &&\mathcal{E}_{14}=J_1+J_2+B, \qquad \mathcal{E}_{15}=-J_1+J_2+B,\nonumber\\
 &&\mathcal{E}_{16}=J_1+J_2+2B.\nonumber
\end{eqnarray}
Where we introduced the following notations
\begin{eqnarray}\label{eq:A2}
 \alpha       &=&\frac{1}{\sqrt{4+2\eta^2}}, \qquad \gamma=\frac{1}{\sqrt{4+2\lambda^2}}, \nonumber\\
 \eta         &=&\frac{(J_1-4J_2) -\sqrt{({J_1}-4{J_2})^2+8d_0^{2}}}{2\,d_0},                      \\
 \lambda      &=&\frac{(J_1-4J_2) +\sqrt{({J_1}-4{J_2})^2+8d_0^{2}}}{2\,d_0},             \nonumber\\
 \lambda \eta &=& -2.                                                                     \nonumber
\end{eqnarray}

\section{\label{sec:appB}Matrix elements $G_{n}$}
The matrix elements $G_6$ and $G_7$ that are used for obtaining Loschmidt echo ($G(t)=|G_6|^2 e^{-i\mathcal{E}_6 t} +|G_7|^2 e^{-i\mathcal{E}_7 t} $) in case of pulse induced dynamics are given as follows
\begin{eqnarray}\label{eq:B1}
 G_{6}&=&\langle\phi_6\vert e^{-i\hat O}\vert\phi_7\rangle                             \\
      &=&4 \alpha\gamma    [        1 - 4 X_1 + 4 X_3 - 2i\lambda(X_2 - X_4)] \nonumber\\
      &+&2 \alpha\gamma\eta[\lambda(1 - 4 X_1 + 4 X_3)- 4i       (X_2 - X_4)] \nonumber\\
      &=&
      - 8i \alpha\gamma(X_2 - X_4)(\eta + \lambda),                           \nonumber\\
 G_{7}&=&\langle\phi_7\vert e^{-i\hat O}\vert\phi_7\rangle \label{eq:B2} \\
      &=&4\gamma^2         [        1 - 4 X_1 + 4 X_3 - 2i\lambda(X_2 - X_4)] \nonumber\\
      &+&2\gamma^2\lambda  [\lambda(1  -4 X_1 + 4 X_3)- 4i       (X_2 - X_4)] \nonumber\\
      &=&2\gamma^2(1 - 4 X_1 + 4 X_3)(2-\lambda^2)
      - 8i\gamma^2\lambda(X_2 - X_4),                                         \nonumber
\end{eqnarray}
where
\begin{eqnarray}\label{eq:B3}
 X_1&=&\frac{1}{8}\left[-        \cos\left(\sqrt{2}d_1\right)+\cosh(\sqrt{2}d_1)     \right],\nonumber\\
 X_2&=&\frac{1}{8}\left[-\sqrt{2}\sin(\sqrt{2}d_1)+\sinh(\sqrt{2}d_1)     \right],\nonumber\\
 X_3&=&\frac{1}{8}\left[         \cos(\sqrt{2}d_1)+\cosh(\sqrt{2}d_1) - 2 \right],  \\
 X_4&=&\frac{1}{8}\left[ \sqrt{2}\sin(\sqrt{2}d_1)+\sinh(\sqrt{2}d_1)     \right].\nonumber
\end{eqnarray}
Substituting (\ref{eq:B3}) in (\ref{eq:B1}) and (\ref{eq:B2}) yields
\begin{eqnarray}\label{eq:B4}
 G_{6}&=&\frac{4i\alpha\gamma(\lambda+\eta)}{\sqrt{2}}\sin(\sqrt{2}d_1),\\
 \label{eq:B5}
 G_{7}&=&\cos(\sqrt{2}d_1)+\frac{8i\gamma^2\lambda}{\sqrt{2}}\sin(\sqrt{2}d_1).
\end{eqnarray}

\end{document}